\documentclass[12pt]{spieman}  

\usepackage{amsmath,amsfonts,amssymb}
\usepackage{graphicx}
\usepackage{setspace}
\usepackage{tocloft}
\usepackage{lineno}

\usepackage[colorlinks=false]{hyperref}

\usepackage{fontawesome5}

\title{Jewel Optics I: non-redundant Fizeau beam combination without the guilt}

\author[a]{Adam K. Taras}
\author[a]{Grace Piroscia}
\author[a, *]{Peter Tuthill}
\affil[a]{Astralis-Usyd, Sydney Institute for Astronomy, School of Physics, University of Sydney, NSW 2006, Australia}

\usepackage{acronym}		
\usepackage{glossaries}		

\usepackage{booktabs}

\newacronym{USyd}{USyd}{the University of Sydney}
\newacronym{ANU}{ANU}{Australia National University}
\newacronym{ESO}{ESO}{European Southern Observatory}

\newacronym{NSF}{NSF}{National Science Foundation}
\newacronym{LIEF}{LIEF}{Linkage Infrastructure, Equipment and Facilities}

\newacronym{VLTI}{VLTI}{Very Large Telescope Interferometer}
\newacronym{ATs}{ATs}{auxiliary telescopes}
\newacronym{UTs}{UTs}{unit telescopes}
\newacronym{ELT}{ELT}{Extremely Large Telescope}
\newacronym{JWST}{JWST}{James Webb Space Telescope}
\newacronym{DCT}{LDT}{Lowell Discovery Telescope}
\newacronym{GMT}{GMT}{Giant Magellan Telescope}
\newacronym{LBT}{LBT}{Large Binocular Telescope}

\newacronym{STS}{STS}{six telescope simulator}
\newacronym{GPAO}{GPAO}{GRAVITY+ adaptive optics}

\newacronym{WFS}{WFS}{wavefront sensor}
\newacronym{ADC}{ADC}{atmospheric dispersion corrector}
\newacronym{LDC}{LDC}{longitudinal dispersion corrector}
\newacronym{AO}{AO}{adaptive optics}
\newacronym{SNR}{SNR}{signal to noise ratio}

\newacronym{OAP}{OAP}{off-axis paraboloid}
\newacronym{DM}{DM}{deformable mirror}
\newacronym{MEMS}{MEMS}{micro-electromechanical system}
\newacronym{IR}{IR}{infrared}

\newacronym{CAD}{CAD}{computer aided design}

\cftpagenumbersoff{figure}
\cftpagenumbersoff{table} 
\begin{document} 
\maketitle

\begin{abstract}
The enduring technique of aperture masking interferometry, now more than 150 years old, is still widely practised today for it opens a window of high angular resolution astronomy that remains difficult to access by any competing technology. However, the requirement to apodise the pupil into a non-redundant array dramatically limits the throughput, typically to $\sim$10\% or less. This in turn has a dramatic impact on sensitivity, limiting observational reach to only bright science targets. This paper presents ``Jewel Optics'', a technology that leverages the gains in signal fidelity conferred by non-redundant Fizeau beam combination without the sensitivity penalty incurred by traditional aperture masks. Our approach fragments the pupil into several sets of sparse-array non-redundant patterns, each of which is encoded onto a unique phase wedge. After extensive searching, solutions could be found where all individual sets are fully non-redundant while fully tiling the available area of the input pupil. Each pattern is assigned a common phase wedge which diverts light from those sub-apertures onto a unique, defined region of the detector. We demonstrate a prototype designed for use in the VAMPIRES instrument at the Subaru telescope and find excellent agreement between the design and lab results. We discuss a design refinement for producing fully achromatic Jewel Optics, and finally we highlight the potential for future work with these optical components. 
\end{abstract}

\keywords{Aperture masking, interferometry, non-redundant masking, high angular resolution}

{\noindent \footnotesize\textbf{*}Peter Tuthill,  \linkable{peter.tuthill@sydney.edu.au} }

\begin{spacing}{1.1}   

\section{Introduction}

Aperture masking interferometry enables astrophysical structures at and somewhat beyond the classical ``diffraction limit'' to be probed, and has established a track record in delivering observables that are robust against atmospheric perturbations\cite{haniff_first_1987}. Despite the genesis of the technique dating back more than 150 years \cite{tuthill_unlikely_2012}, aperture masking continues to play a significant role in high angular resolution astronomy today, enabling recent discoveries from both the ground \cite{han_radiation-driven_2022} and in space \cite{blakely_james_2024}. 

By masking the telescope pupil with a non-redundant pattern of holes, each pair of sub-apertures can be identified with a unique set of interference fringes on the sensor. These fringes are typically studied in the Fourier domain by taking the Fourier transform of the image, enabling the amplitude and phase of these fringes, collectively known as the complex visibility, to be extracted. The van Cittert-Zernike theorem then relates the intensity distribution of the astrophysical source to these visibilities, measured on interferometer baselines defined by the mask. This enables interferometers to uncover high angular resolution structure at scales finer than $\lambda/D$ (where $\lambda$ is the wavelength and $D$ is the diameter of the primary aperture) such as binary companions, the morphology of stellar surfaces and the distribution of circumstellar dust.

\begin{figure}[h]
    \centering
    \includegraphics[width=0.9\textwidth]{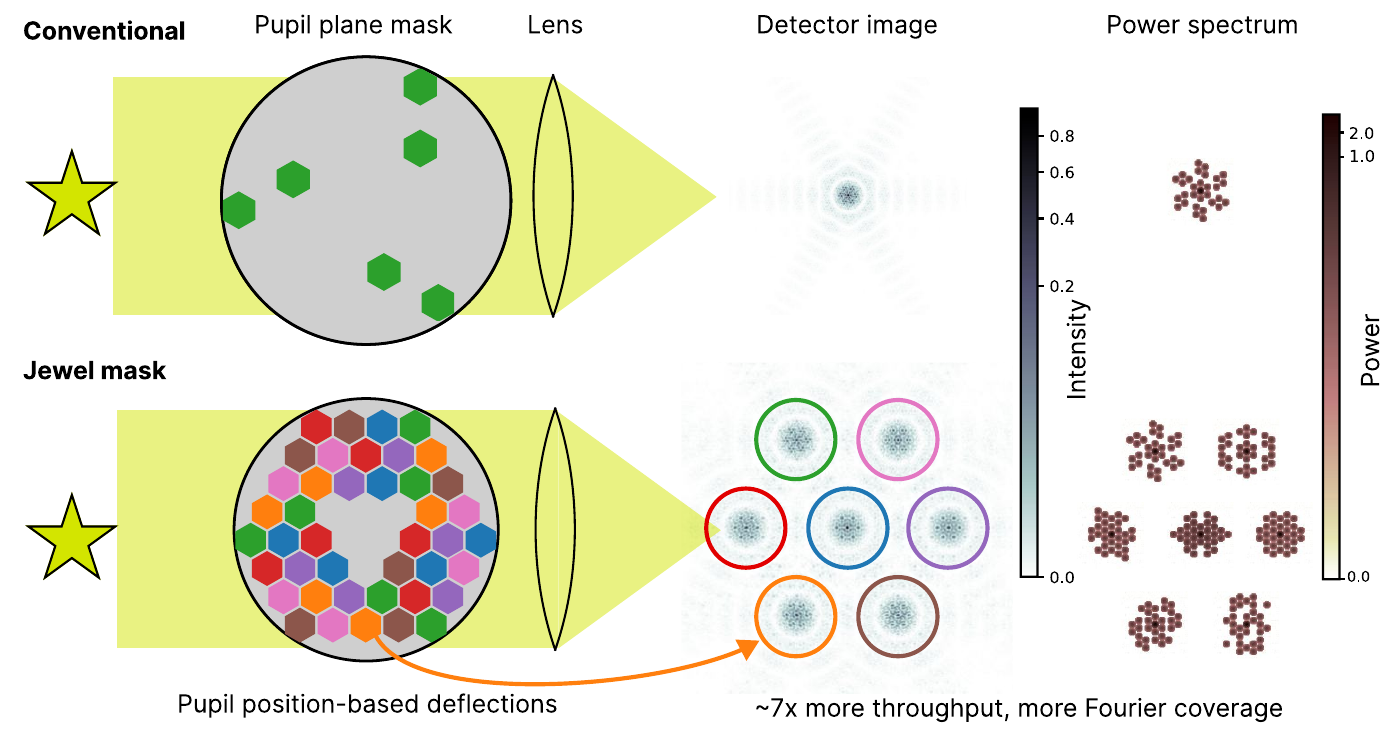}
    \caption{A schematic comparison of conventional aperture masking (top) to the proposed ``Jewel Optics'' proposed here (bottom). In the pupil plane (left), the colour of each Jewel tiling represents a specific phase tilt applied to all same-coloured pattern segments, diverting the incoming wavefront. When performed for each separate pattern, the effect is to produce multiple aperture masks in a single optic, each with its own distinct pointing origin on the sensor. Our approach forms multiple interferograms on the detector (middle), with the interferogram produced by the corresponding pattern of the same colour as circled. Each interferogram has a corresponding power spectrum after cropping (right), revealing dense information about the structure of the astrophysical scene at high angular resolutions. By employing a significantly larger portion of the aperture, Jewel Optics deliver higher sensitivity while preserving the demonstrated advantages of non-redundant Fizeau beam combination.}
    \label{fig:overview}
\end{figure}

In this work, we present Jewel Optics: a novel approach for aperture masking instruments to perform strictly non-redundant beam combination without the major penalty in throughput usually incurred by traditional masks. In principle, arrangements that harness all, or almost all the light collected by the full aperture are possible. The concept is summarised in \autoref{fig:overview}. For a conventional mask, the requirement for non-redundancy of the array demands a drastic reduction in the throughput of the instrument (usually more than 85\% of the pupil is blocked) which in turn has limited application of the technique to bright targets. Jewel Optics overcome this limitation by tessellating multiple non-redundant patterns within a single pupil. Furthermore, Fourier coverage is improved where multiple interferograms are measured simultaneously, improving our ability to extract useful information. Specifically, the key contributions of this work are:
\begin{itemize}
    \item We report sets of tessellations which can fragment the image plane into multiple subarrays, providing more Fourier coverage than a single aperture mask and increasing throughput whilst maintaining the leverage in angular resolution;
    \item We describe manufacturing methods to fabricate Jewel Optics: optical components that redirect designated interferograms onto separate portions of the detector;
    \item We present and evaluate the performance of a simple two-wedge prototype designed for deployment to the Subaru telescope's VAMPIRES instrument, finding excellent agreement between the design specification and the delivered performance; and
    \item We design (in simulation) achromatic Jewel Optics that operate in analogy to achromatic lenses, minimising chromatic aberration and enabling very wide optical bandwidth operation for Jewel Optics at shorter wavelengths operating on large apertures.
\end{itemize}

We first review previous work that aims to increase the efficiency of sparse aperture masks. In doing so, we identify the need for a high throughput, non-redundant aperture mask that can be deployed as a drop-in optic. Next, we develop an algorithm to tessellate the pupil into valid Jewel Optic configurations, searching the space, and finding a total of 14 designs that are published in the appendix of this work. Next, we discuss the strengths and weaknesses of the approach, sketching the region of instrument parameters that this technology is best suited for. We select one design and manufacture a prototype using advanced machining techniques on wedged windows. Through laboratory testing, we validate the performance of the mask prototype and present results, highlighting the agreement with theory. 
We conclude by identifying several exciting avenues of future work developing and exploiting this technology, including on-sky testing and wavefront sensing.

\section{Previous work}
The idea of a lossless fragmentation of a telescope pupil into tessellating non-redundant sub-arrays can be traced to the Keck segment tilting experiment\cite{monnier_mid-infrared_2009}. To accomplish the desired tiling, customised code driving the primary mirror segment alignment system was written to remap the Keck pupil, directing four interferograms onto different sectors on the detector. This enabled the use of 24 of Keck's 36 primary segments, dramatically increasing throughput compared to a conventional aperture mask. Since the primary segments are reflective, this technique was achromatic and could be used over large bandwidths. However, modification to basic observatory infrastructure -- such as real-time codes controlling mirror positions -- is regarded as high risk at observatories, especially in the era of telescopes such as the \gls{JWST} and the \gls{ELT} where optical stability is critical. For segmented telescopes, such architectures are also limited to the grid of existing mirror segments, which provide a lower bound on the size of each sub-aperture. For telescopes with a monolithic primary, tilting cannot be implemented at all. A final reason that no uptake of segment tilting has been reported since the original 15 years ago is that the method is incompatible with adaptive optics systems which are now ubiquitously found feeding high resolution imaging cameras. Jewel Optics, on the other hand, evolve the concept of tilting sub-apertures to a drop-in replacement that fits in a filter wheel, beyond any constraints or limitations inherit to the fragmentation of the pupil at the location of the primary mirror segments themselves.

Holographic aperture masks\cite{tuthill_multiplexed_2018, doelman_first_2021} employ liquid-crystal geometric phase patterns to create multiple interferograms on the detector, also delivering low resolution spectroscopic capabilities. The optic is compact and can be readily introduced into instruments with existing pupil-mask or filter wheels. The capabilities of the holographic components, however, only provide a modest increase to the Fourier coverage beyond that obtained from a single sparse aperture mask. This is because additional interferograms deflected away from the primary on-axis beam are constrained by the highly chromatic manner (holographic gratings) in which the beams are deviated. This, in turn, leads to the requirement that off-axis patterns may only be formed by patterns of sub-apertures that are co-linear. Holographic masking, in common with segment tilting, has seen little activity after initial experiments to prove the concept. 

Finally, pupil remapping methods used in instruments such as DRAGONFLY \cite{robertson_optical_2008,tuthill_photonic_2010,jovanovic_progress_2012, jovanovic_starlight_2012, norris_high-performance_2014} and FIRST \cite{kotani_development_2010, vievard_first_2020,lallement_photonic_2023} employ photonic technologies to transform the two dimensional input pupil into a one dimensional non-redundant fibre array. This grants freedom to add spectral dispersion in an orthogonal dimension on the sensor. This method can, in principle, sample the whole wavefront with beam combination in a non-redundant manner, with spatial filtering at each sub-aperture, and yielding additional wavelength information in the interferogram sampled at high resolution. However, limitations include practical considerations such as sensor area for large, one dimensional patterns and losses when injecting into a single mode fibre. Single mode fibre injection requires a stringent angular tolerance at each sub-aperture whereas Jewel Optics require standard optical tolerances that are readily met by traditional manufacturing methods. A related approach \cite{kim_coherent_2024} interferes the modes harvested by a photonic lantern at the focal plane, effectively inducing complex aperture functions as opposed to the binary valued apertures in conventional masks. Whilst photonic lanterns with larger mode counts could reduce angular tolerances upstream and increase Fourier coverage, there is a quadratic cost in the number of beam combinations needed with mode count. Another iteration of these architectures involves using `oversampled' photonic lanterns \cite{norris_photonic_2024}, where the beam combination is done implicitly in the lantern itself (removing the need for additional beam combination measurements) and then recovered using data driven/machine learning in post processing. These approaches are demanding on observatory infrastructure, requiring a dedicated instrument at an available \gls{AO} corrected focus. They therefore occupy a quite different niche to the simplicity and ease of implementation of aperture masks in existing cameras. 

Our Jewel Optic technology is intended to address shortcomings of prior interferometric approaches. It provides a versatile way to fragment the telescope pupil with few restrictions and limitations on what constitutes a viable pattern. The masks operate as a low-cost, drop-in replacement for a standard-size optic that would occupy a slot in a pupil or filter wheel that are designed for typical locations (including e.g. slowly converging beams) in modern cameras. They deliver a Fizeau beam combination scheme that provides a compact way of sampling visibility information on the detector while leveraging the signal-to-noise benefits conferred by non-redundant pupils.

\section{Discovering Jewel patterns}

Given the intrinsic design freedom to pattern the pupil with arbitrary sub-arrays, we first seek to understand what tilings are possible. We begin by assuming that tiles are placed at vertices of uniform 2-dimensional grid that spans a circular aperture. The number of sets is selected empirically such that the final design has as few sets as possible, with minimal (ideally no) redundancy.
Following ideas originally explored by Golay\cite{golay_point_1971}, we adopt a hexagonal close-packing for these vertices, selecting layouts of a set dimension, removing vertices that clash with existing obstructions such as the secondary mirror. 
However, in some cases, vertices are also removed simply for the purpose of obtaining a search array of the desired dimensionality to effect a good tessellation.

\begin{figure}[h]
    \centering
    \includegraphics[width=0.7\textwidth]{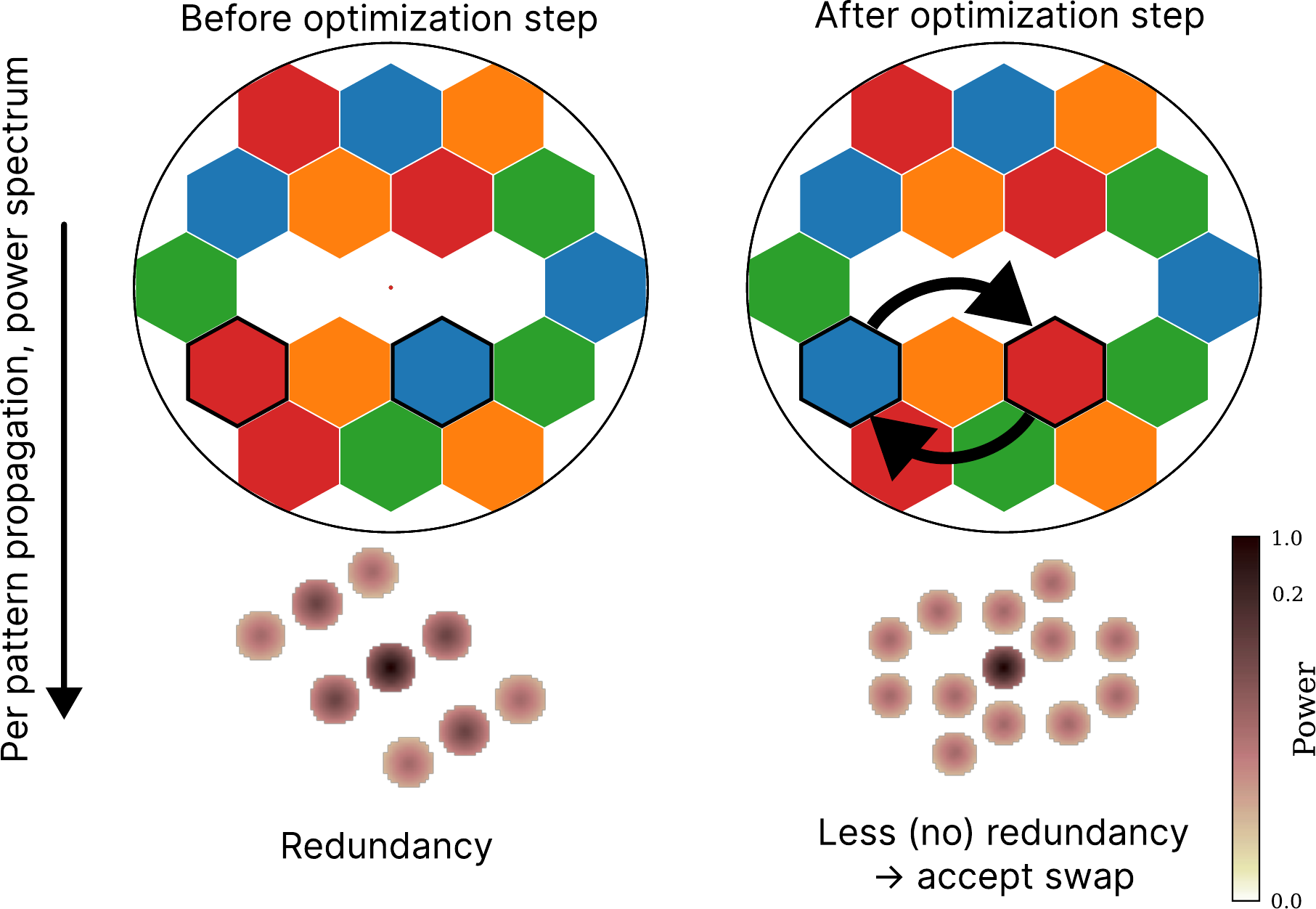}
    \caption{Direct binary search for Jewel pattern discovery attempting to fit 4 patterns (represented with different colours) each with 4 segments into a 16-vertex hexagonal grid. From a valid initial tiling in the pupil plane (top row), the direct binary search algorithm proposes swaps between apertures and accepts changes if the redundancy decreases, measured in the power spectrum for each pattern (blue shown in the bottom row, redundant baselines are where 2 or more power spectral peaks overlap). The resulting tilings inform the way sub-apertures should be deflected onto the detector to generate multiple interferograms, each of which is non-redundant and contains the desired information about the coherence of the astrophysical scene.}
    \label{fig:discover_jewel}
\end{figure}

Over such grids, we discover Jewel tilings using a direct binary search, assigning sets of vertices on the hexagonal grid to different patterns. 
At the start of the algorithm, assignments are done at random. 
The baselines formed by all pairs of sub-apertures in each pattern may then be calculated, and following Golay\cite{golay_point_1971}, determined to be a redundant or non-redundant sampling. 
We adopt the total number of redundant baselines as a merit function to be minimised. A single step of this algorithm is shown in \autoref{fig:discover_jewel}. At each step, a swap between patterns (tile colours in the figure) is proposed, an operation that will modify the power spectrum of the interferograms and hence the redundancy. The algorithm measures the redundancy of the pattern and accepts the swap if the redundancy decreases. We repeat this process, starting from a large number of randomly selected initial states and running until improvements in redundancy slow or saturate, indicating convergence. In the case where multiple solutions beyond permutations and rotations are found, the one with the ensemble of longest baselines is selected, providing the highest angular resolution. 

As a concrete illustration, this work follows the design and manufacture one optimal tiling that contains 4 patterns with 4 sub-apertures (a simple pattern with only a modest count of segments), as shown in the right side of \autoref{fig:discover_jewel}. However, using this same search algorithm we found large numbers of possible patterns with various counts of participating vertices, segments and patterns. Of these we have selected 11 as potential Jewel Optic designs for fitting within circular telescope pupils and spanning a range of sub-aperture sizes, numbers of interferograms and secondary obstruction limits. In addition, we demonstrate the process is useful for telescopes with more complicated pupil geometry, finding two designs for the \gls{LBT} and one for the \gls{GMT}. All designs are summarised and illustrated in Appendix \ref{app:jewel_pattern_spam} and the code used in this work can be found on \href{https://github.com/ataras2/Jewel1}{this GitHub link \faGithubSquare}.

\section{Strengths and weaknesses}

The Jewel Optic solutions we provide will, in general, increase both the throughput and the Fourier coverage when compared to conventional masking. The throughput is improved by a factor roughly equal to the number of interferograms $N_\mathrm{int}$ (assuming each interferogram has the same number of sub-apertures, which is often but not universally the case), ignoring losses at the sub-aperture boundaries and comparing to a mask with the same sub-aperture size. A reasonable number for $N_\mathrm{int}$ from our designs is 4 or 7. The Fourier coverage improvement must be calculated on a case by case basis, however Jewel Optics enable multiple measurements of many spatial frequencies, as shown in \autoref{fig:overview}. In the read-noise limited case, the \gls{SNR} is increased by a factor $\sqrt{N_\mathrm{int}}$, since the total signal is increased by a factor $N_\mathrm{int}$ but it is also spread out over $N_\mathrm{int}$ times more pixels. The \gls{SNR} for baselines measured multiple times will accumulate in the usual way for independent samples.  
A general outcome is that the architecture often provides complete sampling of all baselines within the boundaries of the available sampling grid, usually delivered with non-redundancy, and with several independent and separate measurements of many spatial frequencies simultaneously. In calibration/systematic limited cases, repeatedly sampling a given spatial frequency could be useful to understand wavefront errors, and including an additional term in the direct binary search metric would traverse the tradeoff between depth and breadth of the Fourier sampling.

\begin{figure}[h]
    \centering
    \includegraphics[width=0.5\textwidth]{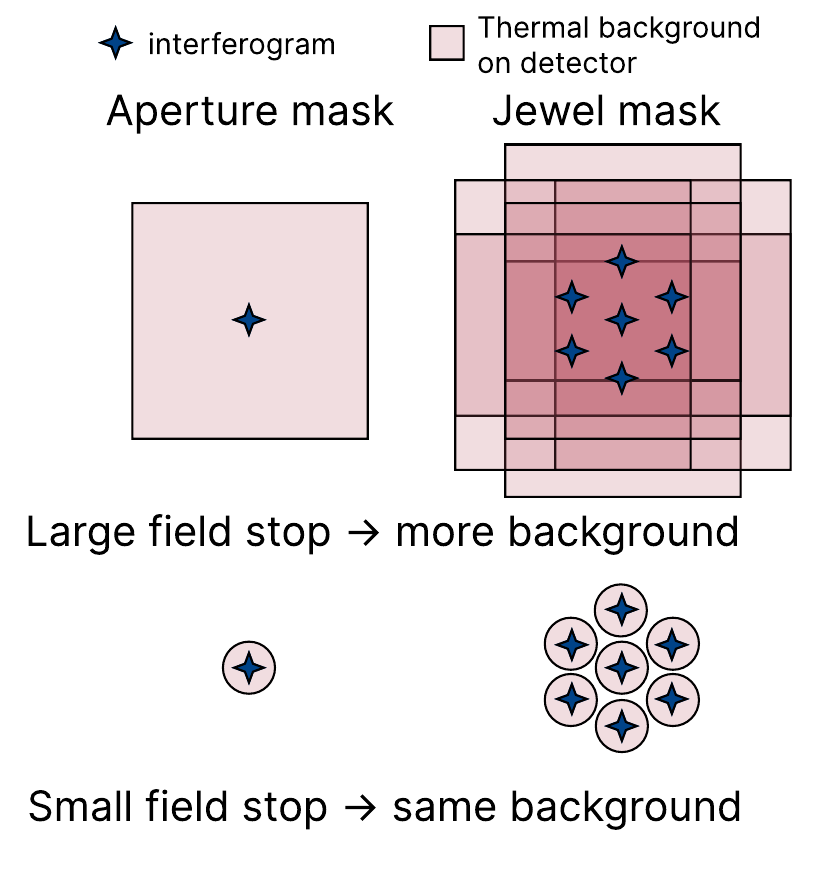}
    \caption{A strategy to ameliorate extra sky background contamination for a Jewel Optic. As opposed to a conventional aperture mask (left column), a Jewel Optic (right column) with a large field stop (top row) would experience $N_\mathrm{int}$ times more sky background. For use cases where noise from the sky background flux sets a significant limitation, the original aperture mask sensitivity may be recovered individually for each Jewel pattern by implementing a tight field stop (bottom row) upstream, sized to match the spacing of the interferograms on the detector. }
    \label{fig:reduce_bg}
\end{figure}

Whilst not relevant to the prototype application developed in this work in the visible bands, we also consider the impact of sky background flux, which can be the limiting case for observations in the thermal infrared. Compared to the case of a classical opaque, perforated sparse aperture mask mounted within a cold wheel in the camera, the background encountered by each interferogram from a Jewel Optic is increased by a factor of $N_\mathrm{int}$, since interferograms still receive background from off-source sub-apertures. This scenario is illustrated in the top row of \autoref{fig:reduce_bg}. Since the signal is spread over $N_\mathrm{int}$ more pixels, the background limited SNR relative to a conventional mask is unity. If, however, a Jewel Optic is designed for a background limited case, then the implementation of a critically-sized upstream field stop can remove all additional background, as shown in the bottom row of \autoref{fig:reduce_bg}. This is typically possible between the relay optics of existing instruments, for example the LBT (see e.g. Spalding \textit{et al.}\cite{spalding_towards_2018}, Figure 3, LMIRcam arm) has multiple filter wheels after the field stop. We note that the resulting background noise level is equal for aperture masks and Jewel pupil remapping. Thus even in the background limited case Jewel Optics may retain the strengths of $N_\mathrm{int}$ times more signal by typically only modifying existing infrastructure, with drop in optic(s) to existing filter wheels.

One obvious drawback of Jewel Optics is a field of view that is limited to at most half the separation between interferograms. The exact scale is a design decision for each particular case, with a trade-off between detector area, chromatic effects, \gls{AO} correction performance and field of view. In almost all instances, however, one can place this at least $10\lambda/D$ away, though it can be much larger; our prototype in this work has a separation of $\sim100\lambda/D$ between interferograms. Science cases for imaging fields larger than this are not at all suited for masking anyway as they lie well beyond the interferometric field of view and therefore require multi-field mosaicing in post-processing. Furthermore, \gls{AO} systems are extremely well suited to resolve structures on these wider scales so they constitute a domain against which masking would not be competitive.

Another consideration that must be dealt with on a case-by-case basis is the effects of the \gls{AO} system, which will generally scatter flux beyond a control radius that depends on the actuator pitch. Interferograms must be placed to clear the region just outside this radius whilst also meeting limitations on detector size and plate scale.

It is our expectation that the most commonly encountered noise floor that will confront Jewel masking is neither background nor photon noise but that which currently limits most aperture masking: systematic optical effects and instabilities that are not well calibrated. Here the complete sampling and non-redundant beam combination offered by the Jewel Optic could deliver further science reach, particularly when deployed in concert with more holistic data reduction strategies that exploit the more complete knowledge of the incident wavefront and optical system delivered. Further exploration of signal-to-noise trades and regimes over which Jewel Optics are competitive will be presented in a forthcoming work.

\section{Manufacturability}

\begin{figure}[h]
    \centering
    \includegraphics[width=0.7\textwidth]{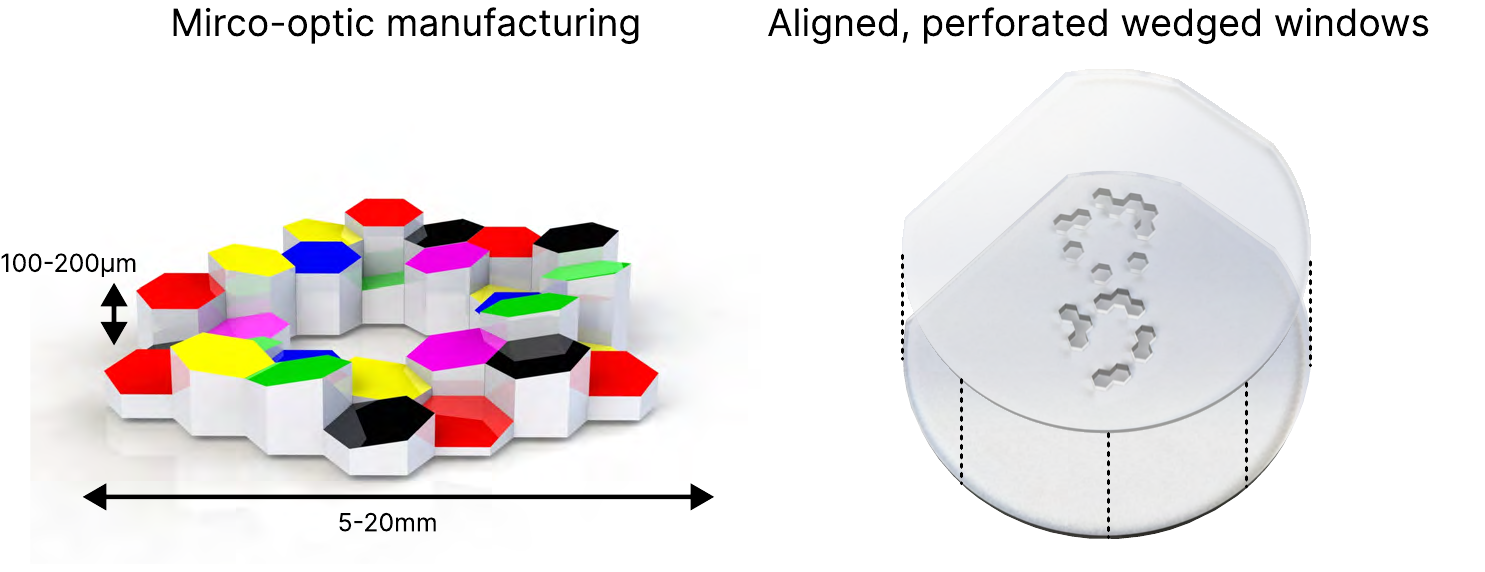}
    \caption{Jewel Optics could be realised using micro-optic additive fabrication (left, vertical dimension oversized by a factor of 10) or stacked wedged windows with custom perforations (right, top wedge made transparent). Typical pupil scales of 5-20\,mm set the assumed manufacturing requirements for each method. }
    \label{fig:manufacturing_options}
\end{figure}

We now consider how to implement the optical pupil fragmentation required by the Jewel Optic designs for the case of a transmissive optic placed in an optical beam train. 
We seek an optic capable of deflecting light passing through specific regions within the pupil. 
This redirection of starlight must be accomplished while maintaining the overall phase coherence of all sub-apertures in that discrete pattern, and would ideally be with minimal or no chromatic dispersion.
In \autoref{fig:manufacturing_options} we illustrate two possible manufacturing techniques that we have employed thus far: (1) micro-optic additive manufacturing and (2) aligned stacks of perforated wedged windows.

For micro-optic additive manufacturing (or, potentially, lithography might also accomplish the required structures), the target design illustrated on the left of \autoref{fig:manufacturing_options} is comprised of sub-aperture structures sculpted such that each unique tiling pattern falls on a distinct plane oriented at an angle governed by the desired angle of deflection. For aperture masking interferometry, the wavefront aberrations between sub-apertures should be much smaller than the wavelength $\lambda$. In our experience, micro-optic fabrication methods were not able to maintain these tolerances over pupil sized (5-20mm) travel, despite several cycles of prototyping and refinement of process. However more specialised machines may be capable of producing acceptable optical components so that additive fabrication of such structures could be explored further.

The second alternative approach mentioned above, employing a stack of perforated wedged windows, has proved capable of yielding acceptable optical quality within available machining technology. Through the use of laser micro-machining or ultrasonic milling, patterns of carefully designed holes can be cut into the wedges, while maintaining the optical quality of the of the substrate in transmission. In the pupil plane, light from each pattern experiences a unique sequence of through-hole (where the light is not deflected) and through-wedge (where the light is deflected by an amount governed by the wedge angle and a direction governed by the wedge orientation), such that each pattern falls onto a unique location of the detector. We convert Jewel patterns into manufacturing drawings by designing sets of holes in windows that uniquely deflect light in all tiles belonging to a given pattern. 
Where many design options are possible to accomplish the same fragmentation, the chosen permutation is that which maximises the number of neighbouring holes in each wedge, minimising the manufacturing time and acting as a heuristic for strength of the overall optic.

Multiple wedged windows are then stacked and oriented in different directions such that the optic deflects light in the desired layout on the sensor. Assembly, alignment and gluing of the final composite optic is performed under a microscope. As each window offers a binary choice -- hole or glass for each sub-aperture -- a stack of $N_w$ windows supports up to $2^{N_w}$ interferograms. Hence, the number of windows needed scales as $\left\lceil \log_2(N_{\mathrm{int}})\right\rceil$ (e.g. the pattern in \autoref{fig:discover_jewel} requires $\left\lceil\log_2(4)\right\rceil=2$ wedged windows). For the case of 2 windows, the angle between the direction of the wedge in the stack is 90$^\circ$, and for 3 windows the angle is $120^\circ$, providing an even separation on the detector in a square or hexagonal grid respectively. To keep within the mechanical thickness allowed by most filter wheel slots (about 5mm) then our typical Jewel Optic will consist of a stack of 2 or 3 wedged windows (maximum 8 interferograms), with a target to manufacture each to be $<$2\,mm thick. Cutting substrates to effect the perforations does not yield ideal edge quality, however imperfections can be masked by manufactured metal shim stencils. We have found that even for the difficult smallest optics, more than 80\% of the available telescope pupil may be sampled, making the throughput gains delivered by the Jewel Optic significant.

Finally, although more technically challenging, a reflective implementation of Jewel Optics comprises an exciting avenue for future work. Manufacturing technologies such as free-form generators \cite{rolland_freeform_2021} or metal sputtering after micro-optic lithography \cite{gonzalezhernandez_microoptics_2023, wang_twophoton_2023} show promise for a future where we can impart a near-arbitrary surface profile at optical precision scales. Manufactured devices would have to meet parallelism tolerances within a fraction of a wavelength for all tiles as well as tip/tilt matching between tiles such that the Airy pattern from each tile overlaps to within a fraction of the first ring. Such a design could also act as a powered optic, for example performing focusing at the same time as pupil fragmentation: a critical flexibility in cases where footprint and throughput are at a premium such as space missions. 
One could also consider segmented deformable mirrors, however for typical cases this is not feasible due to the large travel required to maintain an optically coherent phase wedge. For example, on an 8m telescope with 1.5\,arcsec pattern separation on sky, the required piston extension between the two edges of the pupil is $(8\,\text{m}\times 1.5\,\text{arcsec}/2) = 29\,\mu$m: a factor of a few times larger than any off the shelf device currently provides. We also note that this value is independent of the pupil diameter, so such a system cannot easily be engineered to reduce this limit.

\section{Prototype for Subaru/VAMPIRES}

We present a prototype Jewel Optic fabricated for the VAMPIRES instrument\cite{norris_vampires_2015} at the Subaru telescope, implementing the four-pattern design where each pattern is formed by four sub-apertures, shown on the right of \autoref{fig:discover_jewel}. 
Combined with the recent multi-band improvements\cite{lucas2024visible}, a Jewel Optic in the VAMPIRES instrument would enable high throughput aperture masking interferometry with a grand total of 32 interferograms in a single capture using four wavebands and two polarisations simultaneously. We compute the percentage of light that is incident on the primary (after secondary obstruction) that makes it through the Jewel Optic with losses due to Fresnel reflections, imperfections at the edges of perforations and the limitation of tiling a circular pupil with hexagonal tiles. For a MgF$_2$ Jewel Optic with no anti-reflective coatings, $\pm70\,\mu$m edge defects and air gap between windows (as prototyped in this work), this value is 54\%, which is 3.1 times greater than a single non-redundant mask formed by any one of the individual patterns with the same sub-aperture size. A future optic for the same application could use a Jewel tiling for seven interferograms. Improvements in manufacturing techniques and material selection could reduce edge losses to a few tens of micrometers. In the case of a seven interferogram Jewel Optic with edge losses $\pm20\,\mu$m with anti-reflective coating, the throughput is increased to 69\%, which is 6.5 times larger than a conventional aperture mask with the same sub-aperture size. We also note that the edge losses decrease with increasing pupil size and that the VAMPIRES pupil, at 7\,mm, is particularly small.
Finally we add that future developments may enable index-matched cementing of wedges, reducing the number of surfaces compared to the present air-gapped layers and thereby reducing losses due to Fresnel reflections.

When implementing a Jewel Optic, a critical design parameter is the separation of each interferogram on the detector which, in the wedged window implementation, drives the wedge angle required. For the VAMPIRES design case, we select a separation of 1.5\,arcseconds on sky, which strikes a balance: preventing cross-contamination of the interferograms if they are packed too close while still fitting all 4 images on the very limited field of view (3" square) of the detector.  

With either of the above fabrication methods we must consider the chromatic dispersion. 
In using a glass wedge for its property of deviating the beam, we also inherit the (unwanted) side effect of dispersing the beam in an asymmetric and spectrum dependent way. If the chromatic spread is larger than a fringe, the fringe visibility, and hence the \gls{SNR}, is reduced in a way that cannot be fully recovered in post processing. 
Continuing the illustration with the example of the Jewel Optic for the VAMPIRES instrument, the fringe spacing on the longest baseline ($0.7D$ where $D$ is the diameter of the telescope) in the shortest band (599-652\,nm) is 22\,mas. 
We note that the shortest band is also the one in which the glass has the greatest chromatic effect -- dispersion relations of typical materials rise to shorter wavelengths. In this band, the chromatic dispersion through the required $78\,$arcmin wedge of MgF$_2$ is 3.5\,mas. Repeating the calculation for fused silica (a more standard material) and re-optimising the wedge angle for the desired interferogram separation gives a dispersion of 5.5\,mas. Either of these is smaller than the fringe spacing and so the chromatic smearing of the fringes is acceptably small for either material. 
This could be modelled in post-processing with minimal impact on \gls{SNR}.

A further distinct consideration is that of bandwidth smearing in the $u-v$ plane, where the power spectral peak for a given Fourier baseline will not be isolated at a specific spatial frequency but will smear into a radial line due to the wavelength diversity.
For long baselines and wide bandwidths, bandwidth smearing can cause power from adjacent baselines to overlap (the blue edge of the shorter baseline has the same fringe frequency as the red edge of a longer one). 
This is common to all types of aperture masking, and results in a form of redundancy and aliasing that degrades the signal-to-noise unless accounted for in the optical design to prevent overlapping power.
For the example tiling considered (\autoref{fig:discover_jewel}, right), it is desirable that the power from the 4 unit (in units of sub-apertures) baseline should not bleed into the power from a 5 unit baseline. This imposes a maximum fractional bandwidth of $0.5/4=12.5\,$\%, consistent with the existing complement of VAMPIRES filters.

\section{Results and discussion}

\subsection{Prototype alignment and characterisation}

Custom MgF$_2$ wedged windows that met the required specifications were procured, designed to have a thickness of 1mm at the thick edge. Two optics were procured, with perforations cut using laser micro-machining or ultrasonic milling at the OptoFab and South Australian (SA) nodes of the Australian National Fabrication Facility\footnote{https://anff.org.au/}. For MgF$_2$ substrates, the laser micro-machining produced sharper holes with less surface imperfections and results from this manufacturing process will be used for the remainder of this work.

\begin{figure}[htb]
    \centering
    \includegraphics[width=0.9\textwidth]{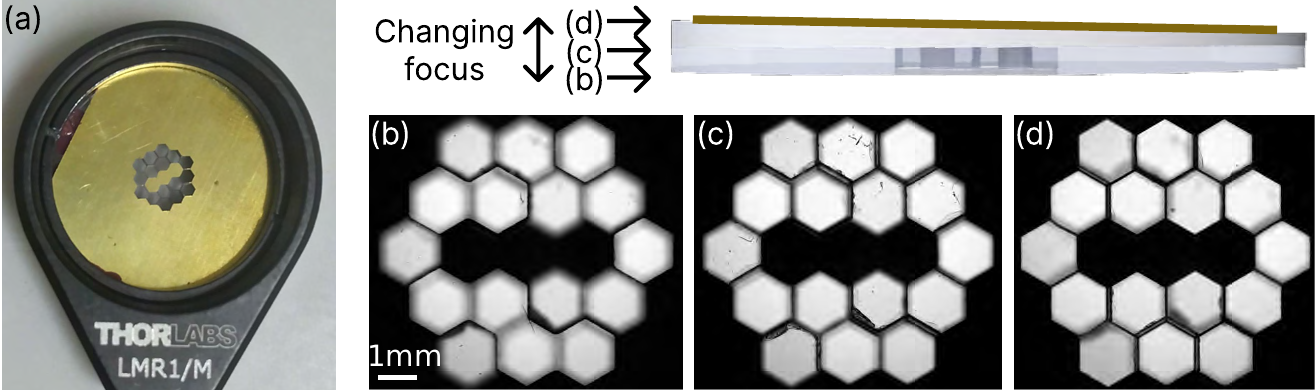}
    \caption{Aligned and glued Jewel Optic. (a) The complete optic has dimension 1" and fits in a standard lens tube or filter mount. Sweeping through the focus (Video, 489KB) on a microscope shows the details of each surface: (b) bottom of the first window including a highlight showing a four sub-aperture hole that was cut, (c) contact between the windows and  (d) the brass shim. Despite some imperfections due to material selection, a near-complete area of each sub-aperture is used.}
    \label{fig:aligned_jewel_masks}
\end{figure}

\autoref{fig:aligned_jewel_masks} shows the assembled prototype Jewel Optic, both as a final assembled optic, and magnified under a microscope. Alignment with a microscope and a slowly setting adhesive produced the best results, with alignment errors on the order of 50\,$\mu$m in a linear offset (measure from the midpoint of the sidewall taper between two wedged windows) and 0.6$^\circ$ in rotation, which is sufficient to not impact the losses from edge effects significantly. Future iterations could use dowel pins beyond the clear aperture to better position multiple windows, which would then be removed after bonding. In this case, we use a colourful nail polish as adhesive making it easy to visually inspect that capillary action hasn't caused the adhesive to obscure the clear aperture. In the microscope images, imperfections are present in regions where the laser cut began, leading to sub-surface stress in the MgF$_2$ substrate. The laser cutting was performed with a 3-axis translation stage. Due to the focus of the cutting laser, an unavoidable sidewall taper of $\sim7^\circ$ is present in the final component. Given the thickness of the substrate, this unwanted sloping surface results in a loss in the region $\pm70\,\mu$m spanning the desired line. An opaque cover shim, made of brass, is designed to block out this region over all sub-apertures, reducing asymmetry in the resulting interferogram. Future manufacturing runs could employ a 5-axis stage for laser cutting, a technology capable of virtually eliminating the sidewall taper. Finally, imperfections remaining on sub-apertures are common to both science and reference star fields and so should be calibrated with typical interferometric observing sequences.

\begin{figure}[htb]
    \centering
    \includegraphics[width=0.7\textwidth]{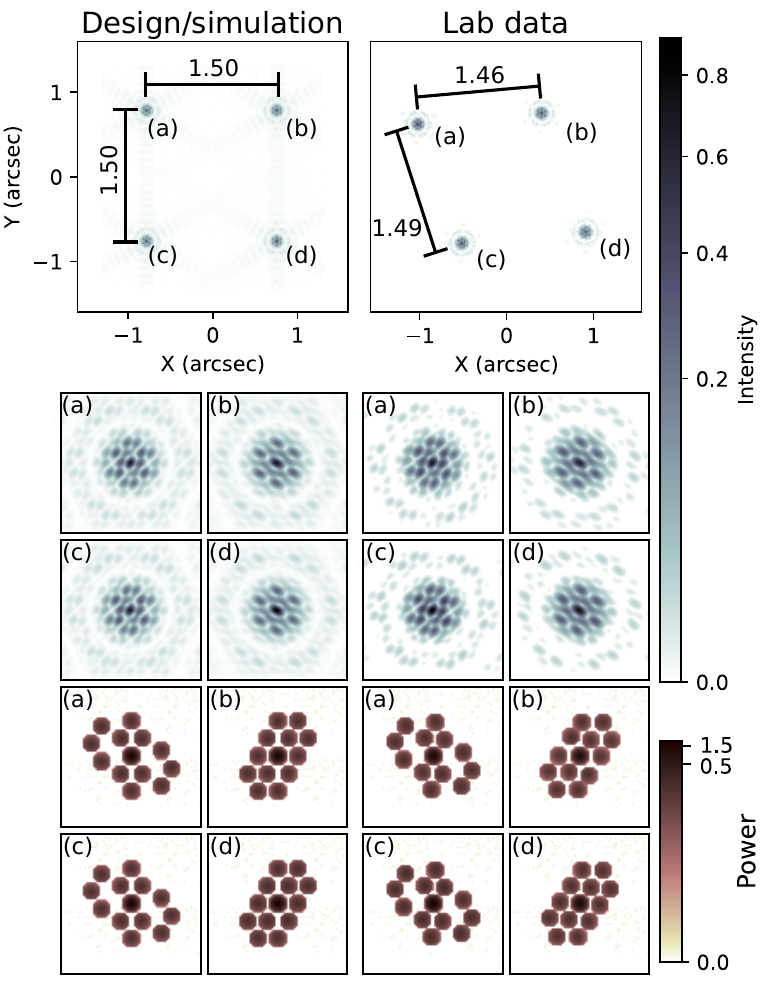}
    \caption{Evaluation of prototype performance of a two-wedge implementation of a Jewel Optic. Simulations (left) and lab data (right) of the optic illuminated in monochromatic light at $\lambda=633$\,nm. All dimensions are in arc-seconds as they would appear on sky at Subaru. The as-designed interferogram pattern (top row, left) matches the designed separation between interferograms well (top row, right). The extracted individual interferograms (middle row) and power spectra (bottom row) also show good agreement with expectations for each pattern, validating the performance of this prototype. }
    \label{fig:results}
\end{figure}

We create a collimated beam from a laser (wavelength $\lambda = 633\,$nm) beam expander that simulates the starlight beam presented to VAMPIRES. After passing through our Jewel Optic an image is formed on a BFS-U3-200S6 sensor\footnote{https://softwareservices.flir.com/BFS-U3-200S6/latest/Model/spec.html} using a $f=400\,$mm lens. After correcting for the non-linearity of the detector response, results are shown in \autoref{fig:results} and exhibit excellent agreement with theory. The interferograms were originally designed to be located on the corners of a square, but instead are seen to form a parallelogram (with angle 75$^\circ$). This has no negative impact for this design, however such manufacturing errors would be detrimental for future designs incorporating a larger number of patterns. 
Upon further investigation, the problem was traced to inaccurate marking of the wedge orientation (typical errors were $\sim7^\circ$) by the manufacturer on the supplied windows. 
For future Jewel Optics, an optical rig has been constructed to accurately re-measure the wedge orientation.

\subsection{Implementing achromatic Jewel Optics}
Whilst not the case for the VAMPIRES instrument, the chromatic dispersion induced by the wedged window could result in a significant smear or blur compared to the diffraction limit. This problem becomes more pronounced for larger apertures and shorter wavelengths (with smaller diffraction limit and generally higher material dispersion). The issue is therefore particularly of interest for the coming generation of \glspl{ELT}. In this section we propose a refinement on the basic wedged windows that incorporate the equivalent of an achromatic doublet Jewel Optics yielding minimal chromatic smearing. A design for deployment to the \gls{LBT} is used as a case study.

\begin{figure}[h]
    \centering
    \includegraphics[width=0.99\textwidth]{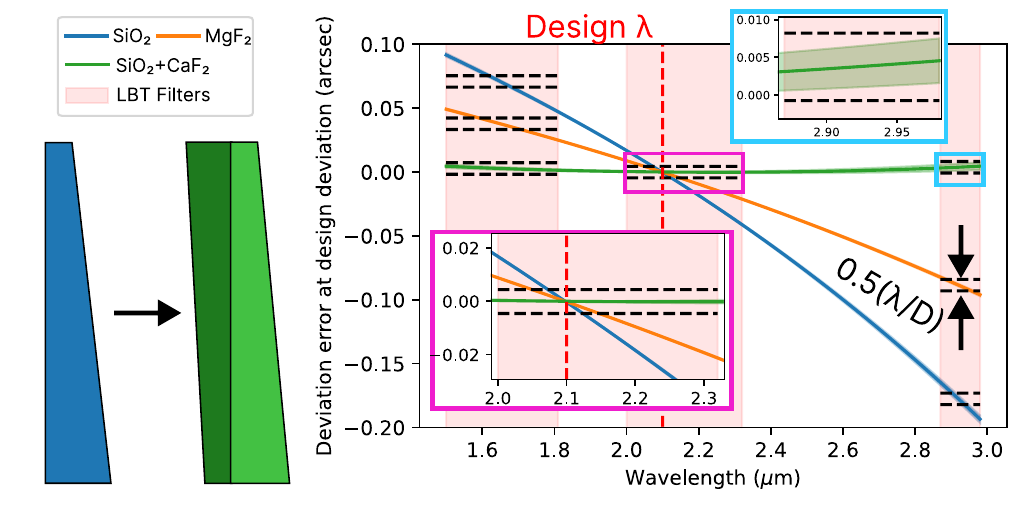}
    \caption{A possible implementation of an achromatic Jewel Optic for the \gls{LBT}. By using two wedged windows with different materials (left), the chromatic dispersion at the detector can be significantly reduced (right). Deviations are measured in arcseconds on sky, after subtracting off the deviation at the design wavelength (2.1\,$\mu$m). The magenta inset focuses on the central filter. Within each filter band the deviation should be reduced to within $\sim\lambda/D$ (black dashed lines), which is challenging at large deviations even for relatively achromatic materials such as magnesium fluoride (MgF$_2$). By using two materials with opposite wedge angles, the dispersion in the central filter is reduced from 30\,mas for MgF$_2$ to 0.2\,mas for the doublet with fused silica and calcium fluoride (CaF$_2$) wedged windows. The plot also includes $3\sigma$ error bars (most visible in the rightmost filter, cyan inset) which show the spread under standard manufacturing tolerances of $\pm5$\,arcmin on each wedged window. These are larger for the achromat design as it employs two windows. However, even the ensemble of samples meets the chromaticity condition when only each instance needs to meet it. }
    \label{fig:achromat_jewel}
\end{figure}

Leveraging insight from lens design, two different materials are employed such that chromatic dispersion cancels. \autoref{fig:achromat_jewel} illustrates this concept and the results for some reasonable materials. When working with multiple materials, we solve a (near trivial, one-dimensional) constrained optimisation problem that minimises the difference in deviation over the whole wavelength range whilst constrained to keep the design deviation at a single wavelength (the red vertical line in \autoref{fig:achromat_jewel}). The dispersive effect is reduced by a factor of 150 in the central band when compared to a single wedge of MgF$_2$, which is among the most intrinsically achromatic optical materials available. To manufacture a single Jewel doublet wedge, the angles would need to be $-305$, $919$\,arcmin for SiO$_2$, CaF$_2$ respectively, versus $697$\,arcmin for a singlet MgF$_2$ wedge alone. Overall this corresponds to an approximate doubling of the final Jewel Optic thickness compared to the simpler wedge designs of the previous sections. We also consider design feasibility respect to manufacturing tolerances, drawing samples from a uniform distribution of the standard tolerance for polishing of $\pm5\,$arcmin. After removing the absolute (wavelength common) effects by considering only the errors relative to the deviation at the design wavelength, we see that even the ensemble of designs fits within the chromaticity constraints (cyan inset in \autoref{fig:achromat_jewel}), and hence any individual sample readily meets this criterion.

Whilst our analysis here has demonstrated that achromatic doublet Jewel Optics are feasible and beneficial, future work could further investigate optimisation that is cognisant of both wedge angle and misalignment between wedged windows of the same pair, as well as transmission considerations and hard thickness constraints. Solving this optimisation problem for all combinations of materials would be tractable for example with Monte Carlo simulations deploying gradient descent on a linear combination of loss functions encoding the chromatic effects and throughput, as well as results from further manufacturing experiments on edge quality for different materials.

\section{Conclusions and future work}

In this work we present Jewel Optics: optics that interlace multiple non-redundant aperture masks in a pupil plane to create interferograms with complete Fourier coverage and high throughput compared with conventional aperture masks. This develops ideas originating with the Keck segment tilting experiment but allowing tilings not limited by primary mirror segments, implemented instead with a drop-in optic, deployable in  almost any high resolution imaging camera. We present sets of tilings for various apertures and samplings obtained through a direct binary search, where patterns are optimised to contain non-redundant pupil sub-arrays. In total we document 11 possible masks for circular apertures, two masks for the \gls{LBT} pupil and one for the \gls{GMT}. We explore hardware implementations to create such an optic, presenting the manufacturing and lab testing of a two wedge prototype designed for the VAMPIRES instrument. We find that our approach effectively generates multiple simultaneous aperture masks with excellent optical quality, increasing throughput by a factor 3.1 on this prototype compared with a conventional aperture mask with the same sub-aperture size, and measuring significantly more Fourier coverage when compared to a similarly sized aperture mask. We envision further improvements in manufacturing to enable masks with smaller holes, where these techniques could feasibly reach 69\% throughput (6.5 times larger than a conventional mask). Finally, we show how our approach could be made achromatic with a pair of wedges in a way analogous to achromatic doublet lenses. This refinement would open the way for implementations to shorter wavelengths and/or very large apertures of coming generation observatories.

This novel implementation of the enduring technique of aperture masking provides a means of extending throughput and Fourier coverage of existing instruments, ultimately enabling deeper exploration of fainter targets. We are also currently investigating Jewel Optics for the \gls{DCT}, \gls{LBT}, Keck telescope and the \gls{GMT}, believing there is a clear path for future work with detailed lab characterisation to fit the imperfections of the mask and subsequent on-sky testing of these optics. A \gls{DCT} mask could be compared to previously successful speckle techniques in the visible on smaller telescopes, while demonstrations on the \gls{LBT} would be an important precursor for \gls{ELT} instruments and are a natural extension of previous science cases for aperture masking \cite{tuthill_unlikely_2012}. Furthermore, since non-redundant beam combination facilitates a unique mapping between fringe information extracted from interferograms and the phasing of the pupil segments from which the light originates, Jewel Optics therefore encode information about wavefront phasing. This dual promise as simultaneous science imager and a wavefront sensor also has the advantage that the wavefront measurements are made at the location of the science sensor eliminating non-common path errors. Information from Jewel Optic wavefront sensing may therefore complement local integrating wavefront sensors (e.g. Shack–Hartmann) that are insensitive to discontinuities such as petaling modes, as well as to better inform post-processing methods.

\appendix    

\section{Jewel tilings over many geometries}
\label{app:jewel_pattern_spam}

\subsection{Tessellations for circular aperture telescopes }

\autoref{tab:circ_jewel} summarises the key properties of the Jewel Optic tessellations found on grids fit within circular apertures. These comprise the best performing patterns of many thousands of candidates recovered, selected for lowest redundancy, Fourier coverage and spanning a range in tilings intended to give flexibility for various use cases.
They are ordered from least to most complex, featuring anywhere between 4-9 interferograms. The first column contains a clickable reference to the figure showing the mask. The underlying grid could be situated within the circular pupil in two ways: with the exact centre of the circle occupied by a vertex in the tile pattern, or by the centre of the face of a tile. 
This choice is listed in the second column. 
The other columns give the properties of each pattern, as described in the table itself.

The figures following show the Jewel Optic in the pupil plane on the left and each power spectrum separated on the right, with a central coloured dot matching the power spectrum to the tiling colour. Redundant baselines, if any, are highlighted in the power spectrum as a patch of different colour where the splodges overlap creating redundancy. 
The pattern depicted in \autoref{fig:30-segment Centered grid} is the same as that previously used in the Keck segment tilting experiment\cite{monnier_mid-infrared_2009}, although only 4 patterns of 6 segments were used, omitting the additional two triangles of 3.

\begin{table}[H]
\centering

\caption{A summary of selected Jewel patterns suitable for circular aperture telescopes. Each row details a different mask design. Columns describe: figure number, the centring of the grid used (face or vertex centred), the incircle diameter of the sub-aperture $d_\mathrm{sub-ap}$, the total number of interferograms $N_{\mathrm{int}}$, the number of sub-apertures per interferogram $N_\mathrm{sub-ap}/N_\mathrm{int}$, the number of baselines $N_\mathrm{base}$, the number of redundant baselines $N_\mathrm{red}$, the maximum possible diameter of the secondary $\max D_s$, and the maximum baseline present $\max |B|$. Arrows in the column headers indicate the desired direction for the parameter. In the case where the Jewel Optic has different numbers of sub-apertures per interferogram, the count is separated with a `+' and `,' i.e. 3+1\,$ | $\, 5,4 indicates 3 interferograms with 5 sub apertures, and 1 interferogram with 4 sub apertures.}
\label{tab:circ_jewel}
\begin{tabular}{ccccccccc}
\toprule
Fig & Centre & $d_\mathrm{sub-ap}$ & $N_{\mathrm{int}}$ & $N_\mathrm{sub-ap}/N_\mathrm{int}$ & $N_\mathrm{base}\,(\uparrow)$ & $N_\mathrm{red}\,(\downarrow)$ & $\max D_s $  & $\max |B|\,(\uparrow)$  \\ \midrule

 \ref{fig:FC4x4s} & Face & 0.196 & 4 & 4 & 24 & 0 & 0.196  &  0.707  \\

 \ref{fig:VC4x5s} & Vertex & 0.166 & 3+1 & 5,4 & 40 & 6$^\dagger$ & 0.332  &  0.760    \\

 \ref{fig:30-segment Centered grid} & Face & 0.141 & 4+2 & 6,3 & 66 & 0 &  0.327 &  0.648  \\

 \ref{fig:36-segment 7-pattern NR solution} & Face & 0.141 & 6+1 & 5,6 & 75 & 0 &  0.141 &  0.748 \\

\ref{fig:VC7x6s}  & Vertex & 0.125 & 7 & 6 & 105 & 0 & 0.249  &  0.758 \\

 \ref{fig:FC7x6s_42} & Face & 0.110 & 7 & 6 & 105 & 0 &  0.460 & 0.883 \\

 \ref{fig:6 sets of 7 segment arrays} & Face & 0.110 & 6 & 7 & 126 & 10$^\dagger$ &  0.460 &  0.834 \\

 \ref{fig:48-segment Vertex-Centered Grid} & Face & 0.110 & 6+1 & 7,6 & 141 & 24$^\dagger$ &  0.331 & 0.883 \\

\ref{fig:FC4x7s+3x6s_46}  & Face & 0.110 & 4+3 & 7,6 & 129 & 0 &  0.331 &  0.834 \\

 \ref{fig:FC4x7s+4x6s_52} & Face & 0.110 & 4+4 & 8,7 & 144 & 0 &  0.255 & 0.834  \\

 \ref{fig:9x6 fully NR pattern} & Face & 0.110 & 9 & 6 & 135 & 0 &  0.255 &   0.883 \\\bottomrule
 \multicolumn{9}{l}{$^\dagger$ Not non-redundant: better solutions may exist but our algorithm did not find any} \\
 \multicolumn{9}{l}{after extensive searching.}\\
\end{tabular}
\end{table}

\begin{figure}[H]
    \centering
    \includegraphics[width=0.712\textwidth]{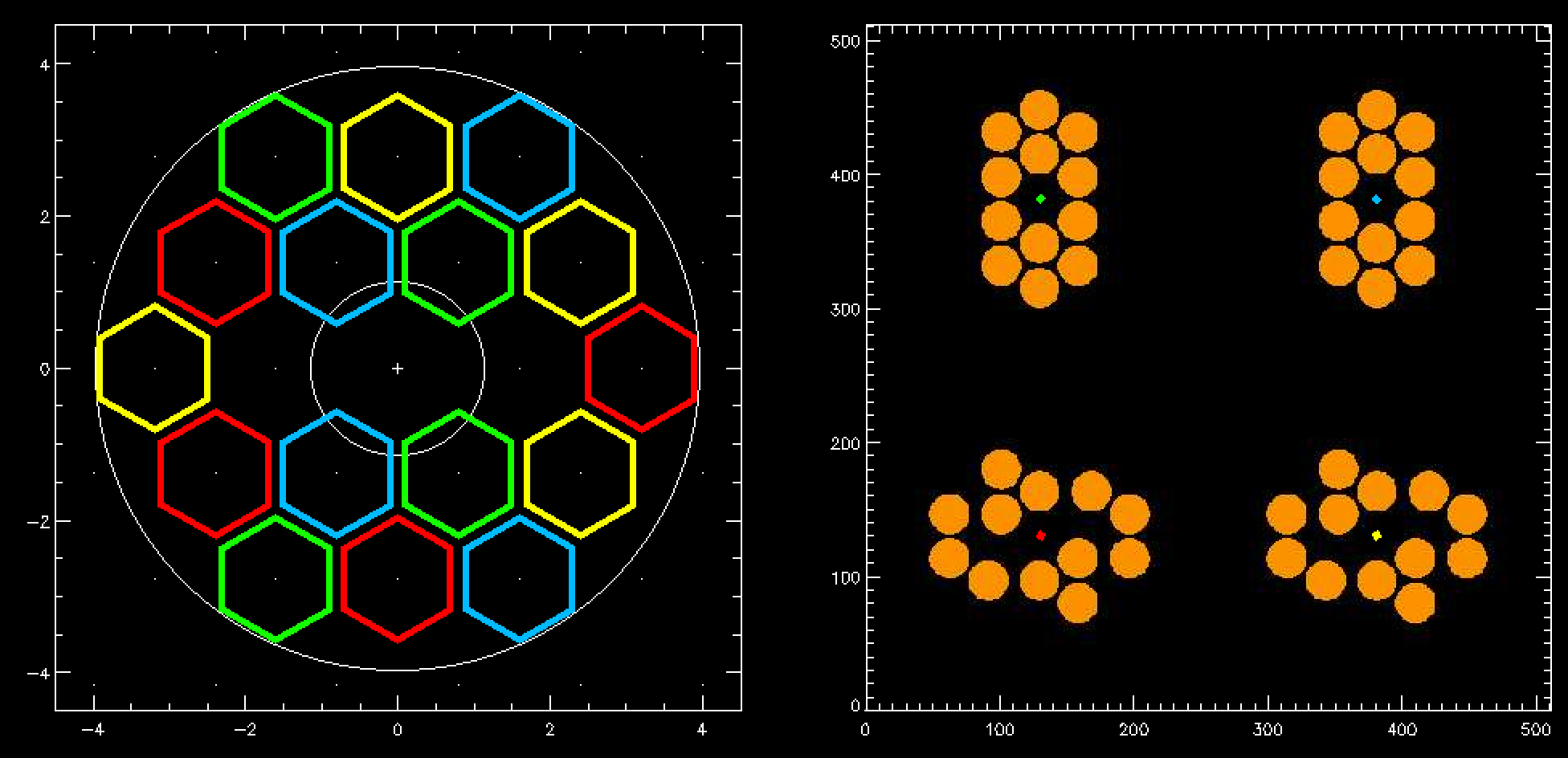}
    \caption{Left panel: Hypothetical circular pupil with central circular secondary mirror obstruction. Overlaid coloured hexagons identify $N_\mathrm{int}$ patterns that tessellate the available hexagonal close packed grid of vertices. Right panel: Fourier coverage of each of the $N_\mathrm{int}$ interferograms. }
    \label{fig:FC4x4s}
\end{figure}

\begin{figure}[H]
    \centering
    \includegraphics[width=0.712\textwidth]{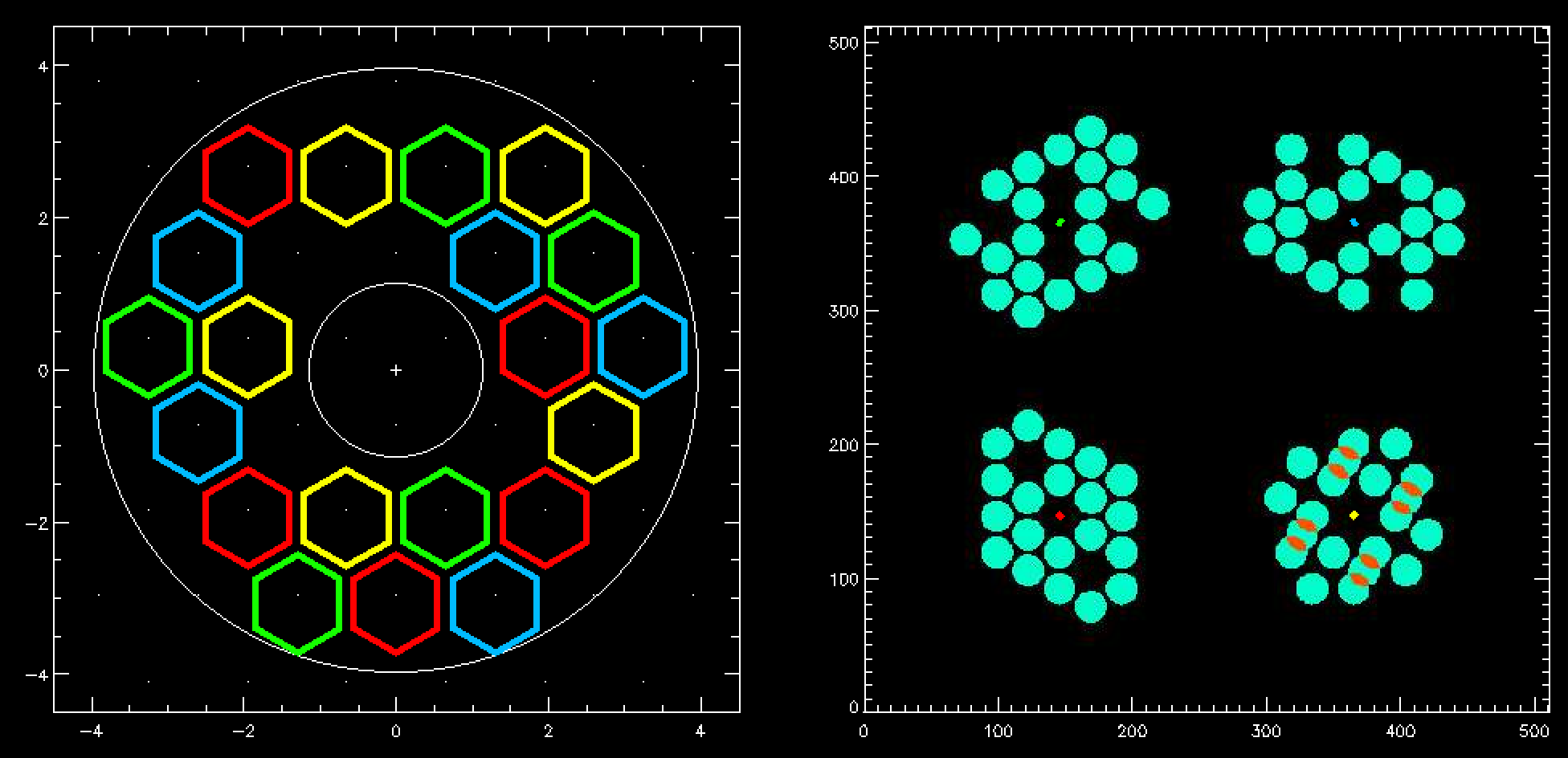}
    \caption{See caption for \autoref{fig:FC4x4s}.}
    \label{fig:VC4x5s}
\end{figure}

\begin{figure}[H]
    \centering
    \includegraphics[width=0.712\textwidth]{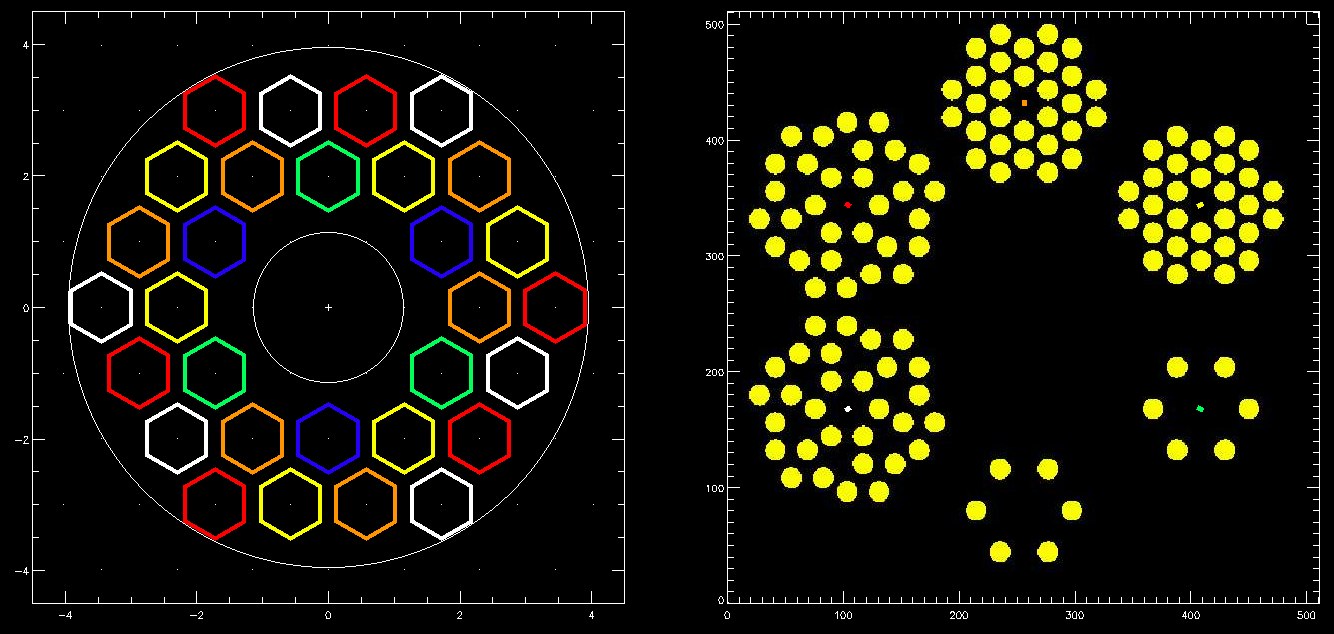}
    \caption{See caption for \autoref{fig:FC4x4s}.}
    \label{fig:30-segment Centered grid}
\end{figure}

\begin{figure}[H]
    \centering
    \includegraphics[width=0.712\textwidth]{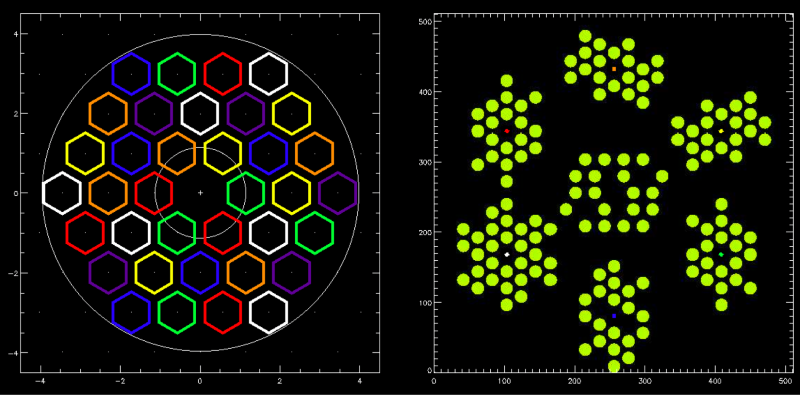}
    \caption{See caption for \autoref{fig:FC4x4s}.}
    \label{fig:36-segment 7-pattern NR solution}
\end{figure}

\begin{figure}[H]
    \centering
    \includegraphics[width=0.712\textwidth]{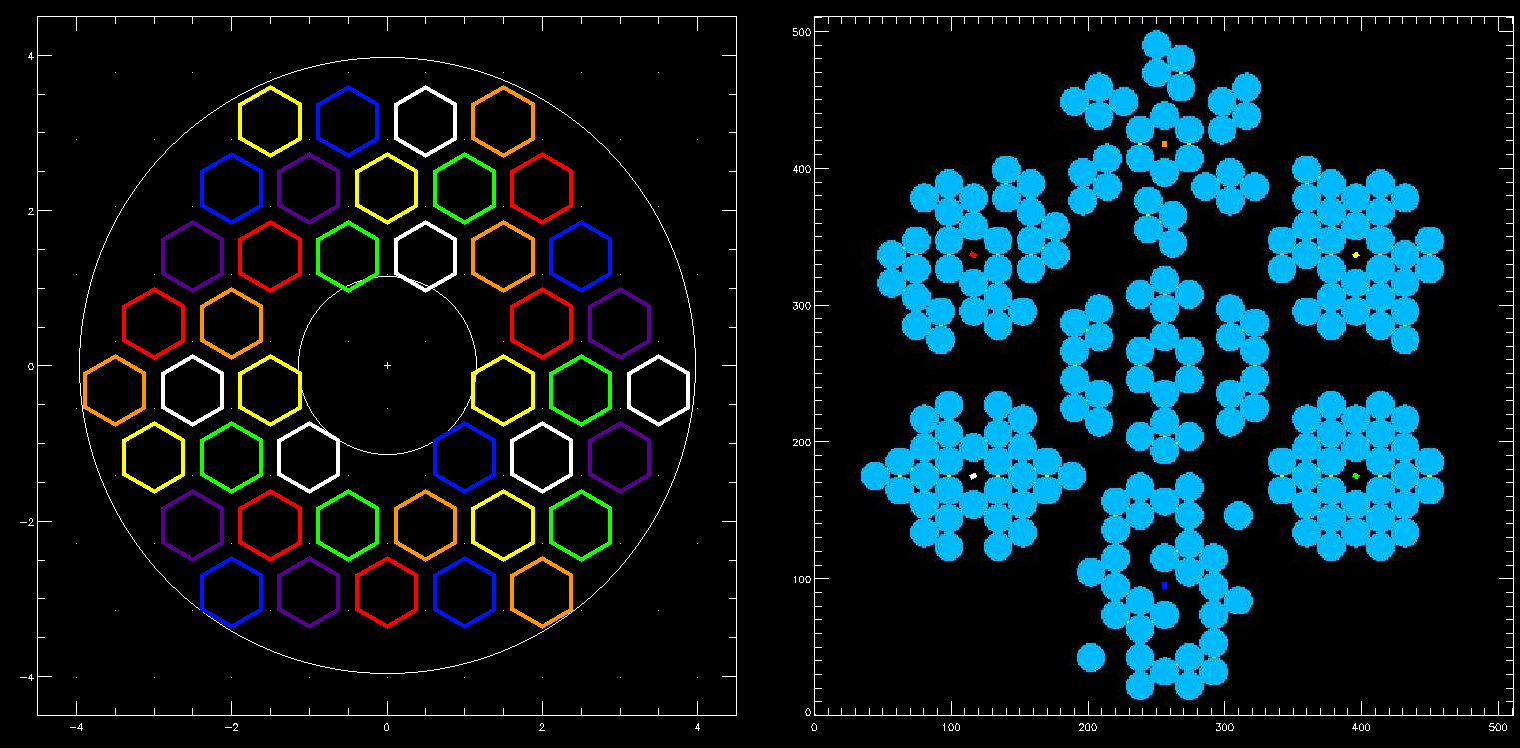}
    \caption{See caption for \autoref{fig:FC4x4s}.}
    \label{fig:VC7x6s}
\end{figure}

\begin{figure}[H]
    \centering
    \includegraphics[width=0.712\textwidth]{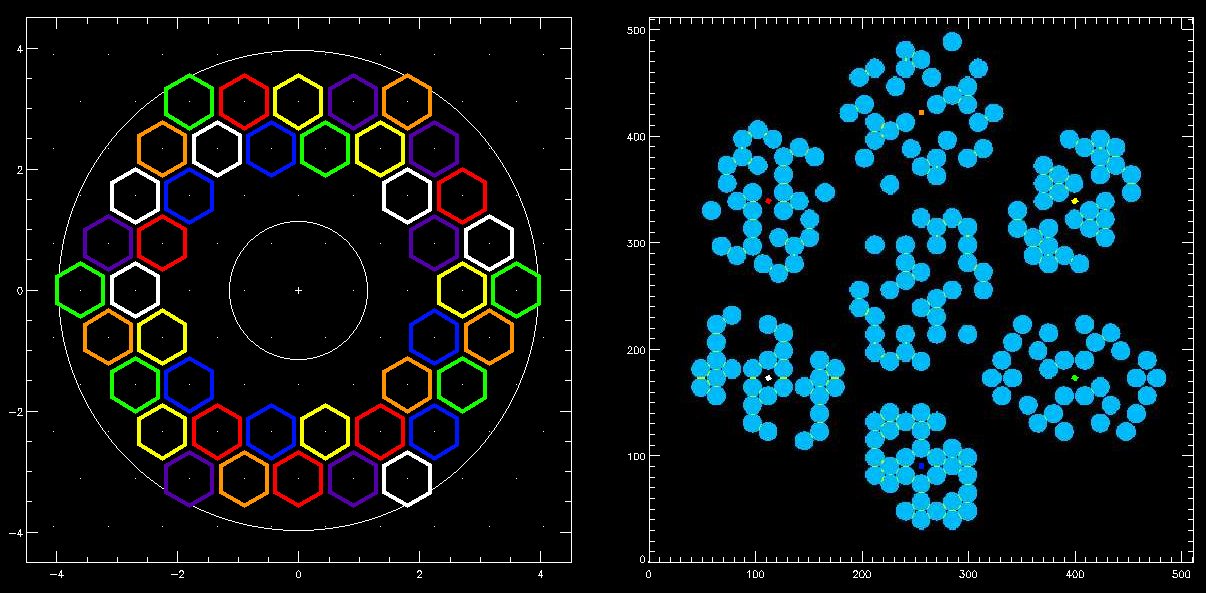}
    \caption{See caption for \autoref{fig:FC4x4s}.}
    \label{fig:FC7x6s_42}
\end{figure}

\begin{figure}[H]
    \centering
    \includegraphics[width=0.712\textwidth]{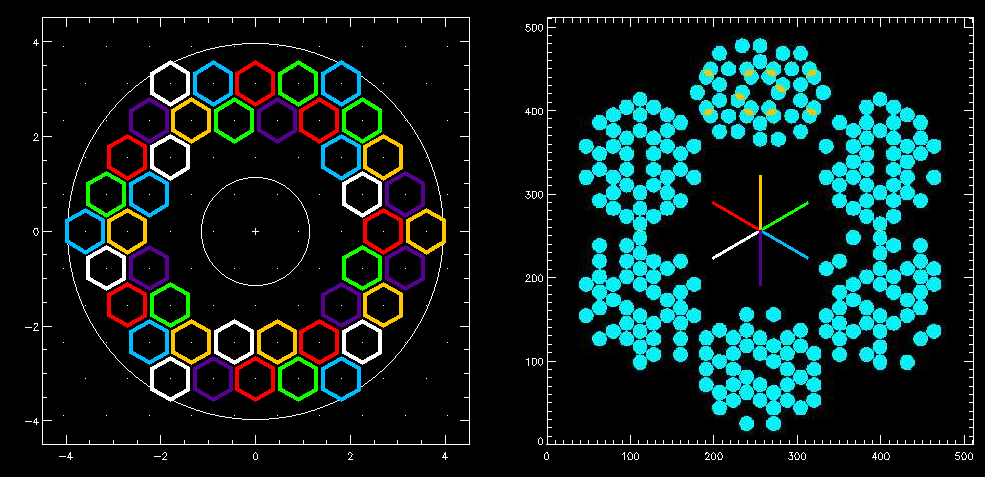}
    \caption{See caption for \autoref{fig:FC4x4s}.}
    \label{fig:6 sets of 7 segment arrays}
\end{figure}

\begin{figure}[H]
    \centering
    \includegraphics[width=0.712\textwidth]{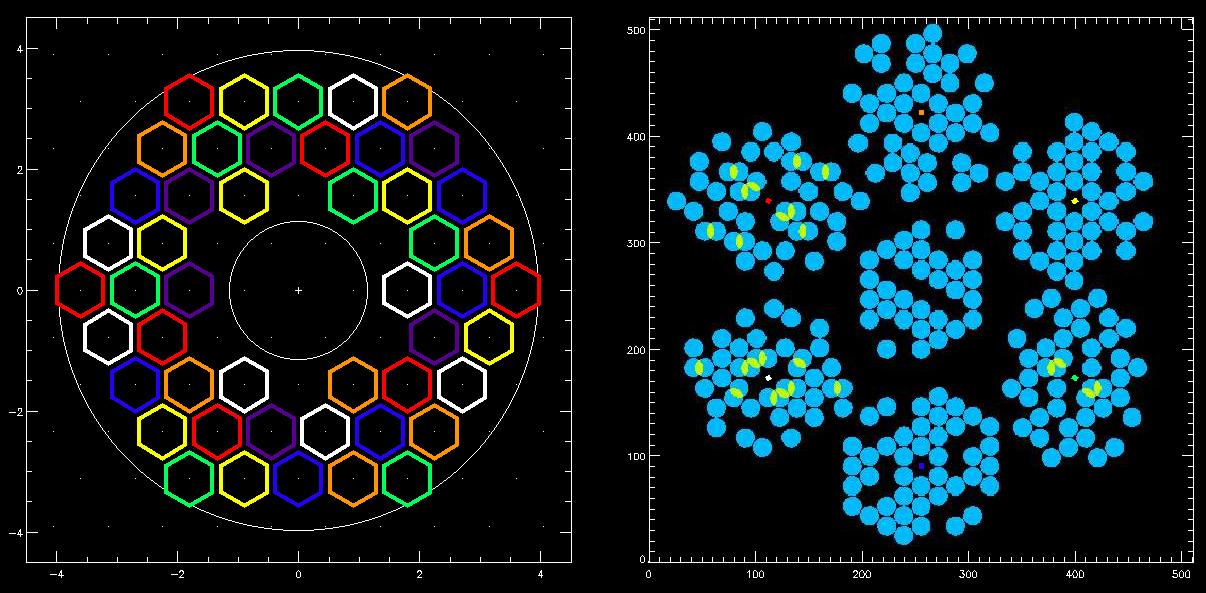}
    \caption{See caption for \autoref{fig:FC4x4s}.}
    \label{fig:48-segment Vertex-Centered Grid}
\end{figure}

\begin{figure}[H]
    \centering
    \includegraphics[width=0.712\textwidth]{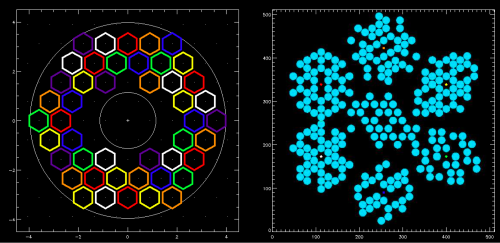}
    \caption{See caption for \autoref{fig:FC4x4s}.}
    \label{fig:FC4x7s+3x6s_46}
\end{figure}

\begin{figure}[H]
    \centering
    \includegraphics[width=0.712\textwidth]{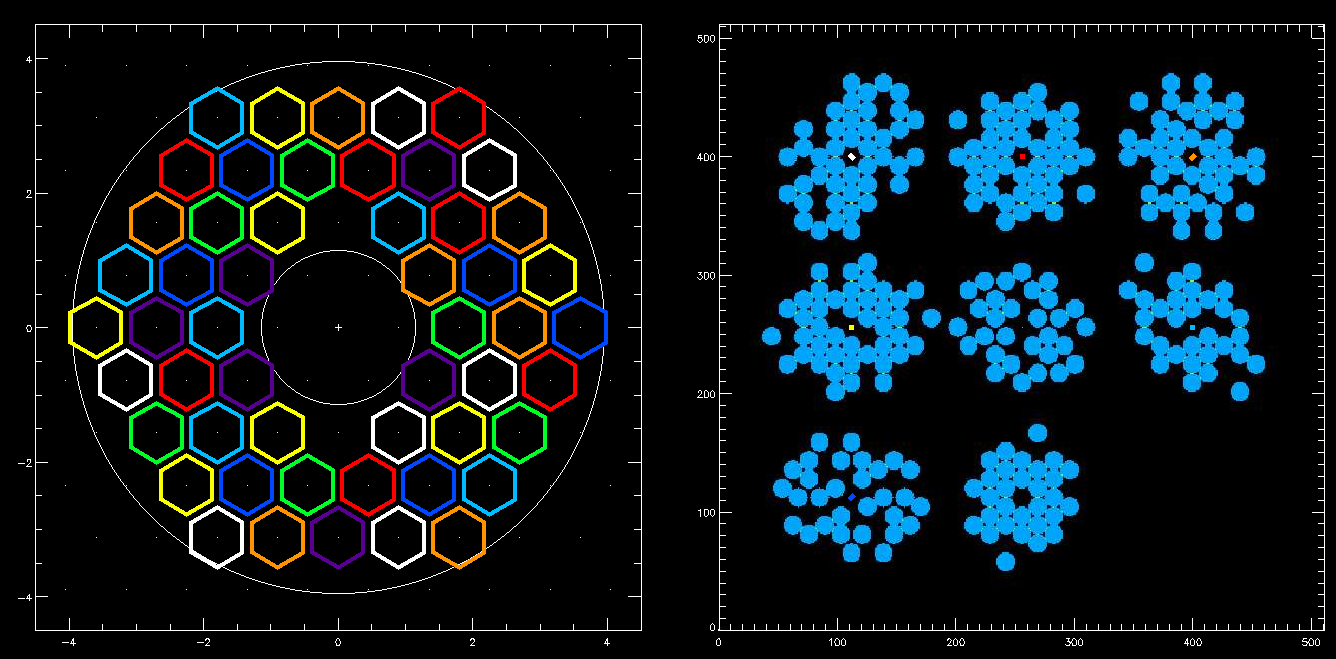}
    \caption{See caption for \autoref{fig:FC4x4s}.}
    \label{fig:FC4x7s+4x6s_52}
\end{figure}

\begin{figure}[H]
    \centering
    \includegraphics[width=0.712\textwidth]{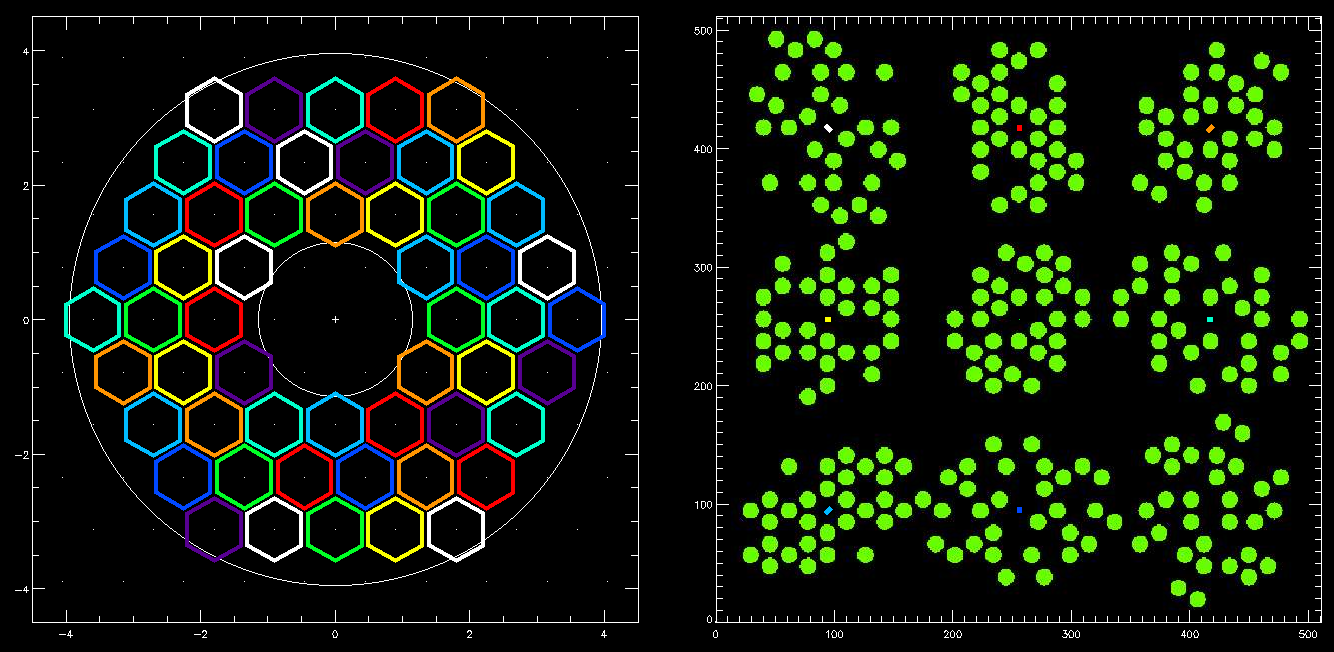}
    \caption{See caption for \autoref{fig:FC4x4s}.}
    \label{fig:9x6 fully NR pattern}
\end{figure}

\subsection{Tessellations for multi-aperture telescopes}

\autoref{tab:LBT_GMT_solns} similarly shows a summary for two multi-aperture telescopes: the \gls{LBT} and \gls{GMT}. Dimensions are now given in physical units since the telescopes are fixed. 

\begin{table}[H]
\centering
\label{tab:LBT_GMT_solns}
\caption{A summary of the two solutions for the Large Binocular Telescope and one for the Giant Magellan Telescope. All lengths are n ow in meters since the mirror size is fixed. The other columns are as described in \autoref{tab:circ_jewel}.}
\begin{tabular}{ccccccccc}
\toprule
Telescope &Fig & Centre & $d_\mathrm{sub-ap}$ (m) & $N_{\mathrm{int}}$ & $N_\mathrm{sub-ap}/N_\mathrm{int}$ & $N_\mathrm{base}$ & $N_\mathrm{red}$ &  $\max |B|$ (m)\\ \midrule

LBT&\ref{fig:LBT6x6s_36}  & Face & 1.618 & 6 & 6 & 90 & 0 & 19.617  \\

LBT& \ref{fig:LBT6x7s+1x6s_48} & Vertex & 1.369 & 6,1 & 7,6 & 141 & 0 &  18.414 \\

GMT&\ref{fig:GMT8x6s_48}  & Face & 2.609 & 8 & 6 & 120 & 0 &  18.813 \\
\bottomrule

\end{tabular}
\end{table}

\begin{figure}[H]
    \centering
    \includegraphics[width=0.712\textwidth]{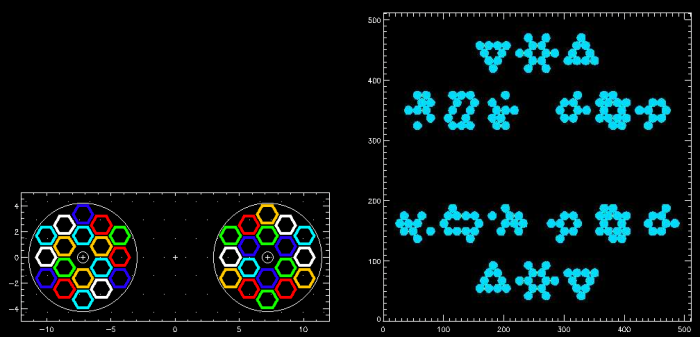}
    \caption{See caption for \autoref{fig:FC4x4s}.}
    \label{fig:LBT6x6s_36}
\end{figure}

\begin{figure}[H]
    \centering
    \includegraphics[width=0.712\textwidth]{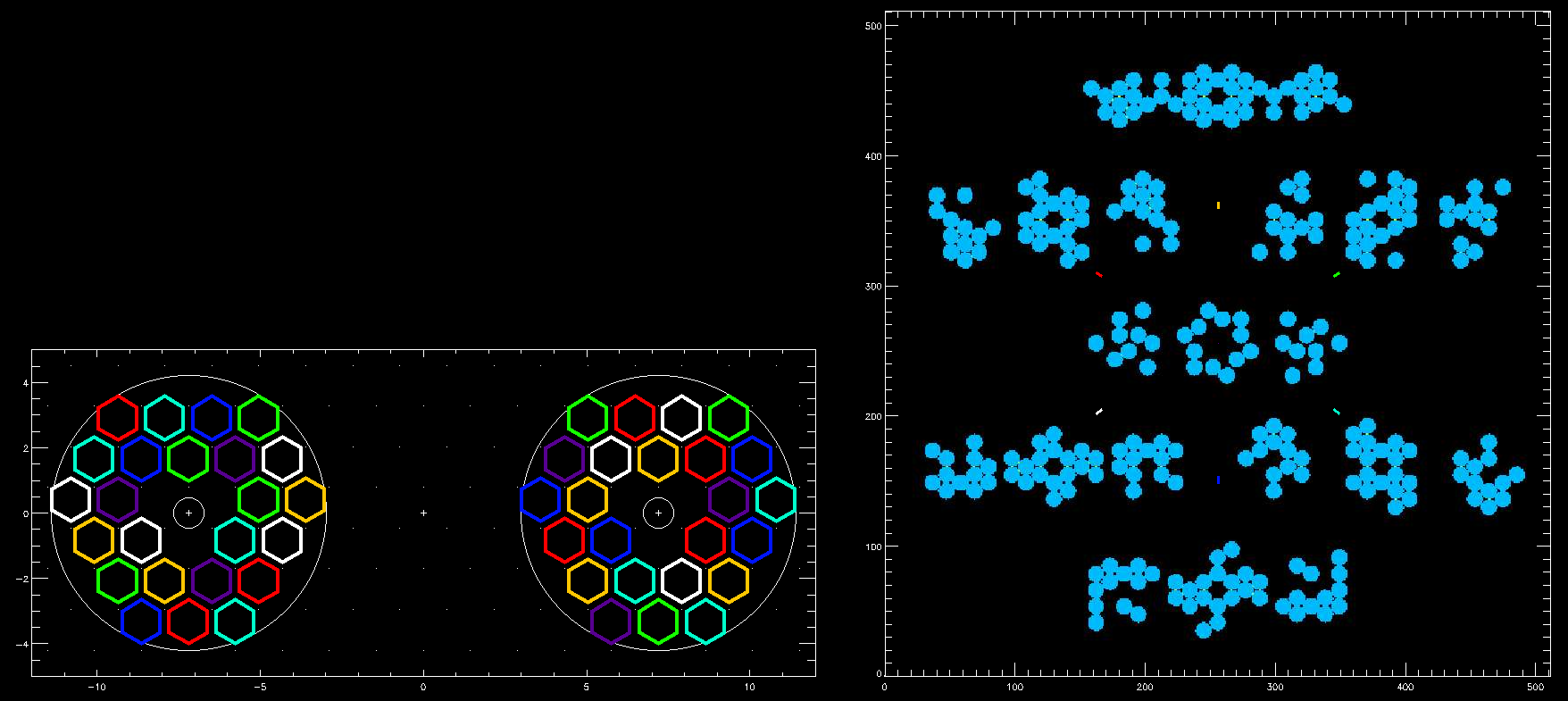}
    \caption{See caption for \autoref{fig:FC4x4s}.}
    \label{fig:LBT6x7s+1x6s_48}
\end{figure}

\begin{figure}[H]
    \centering
    \includegraphics[width=0.712\textwidth]{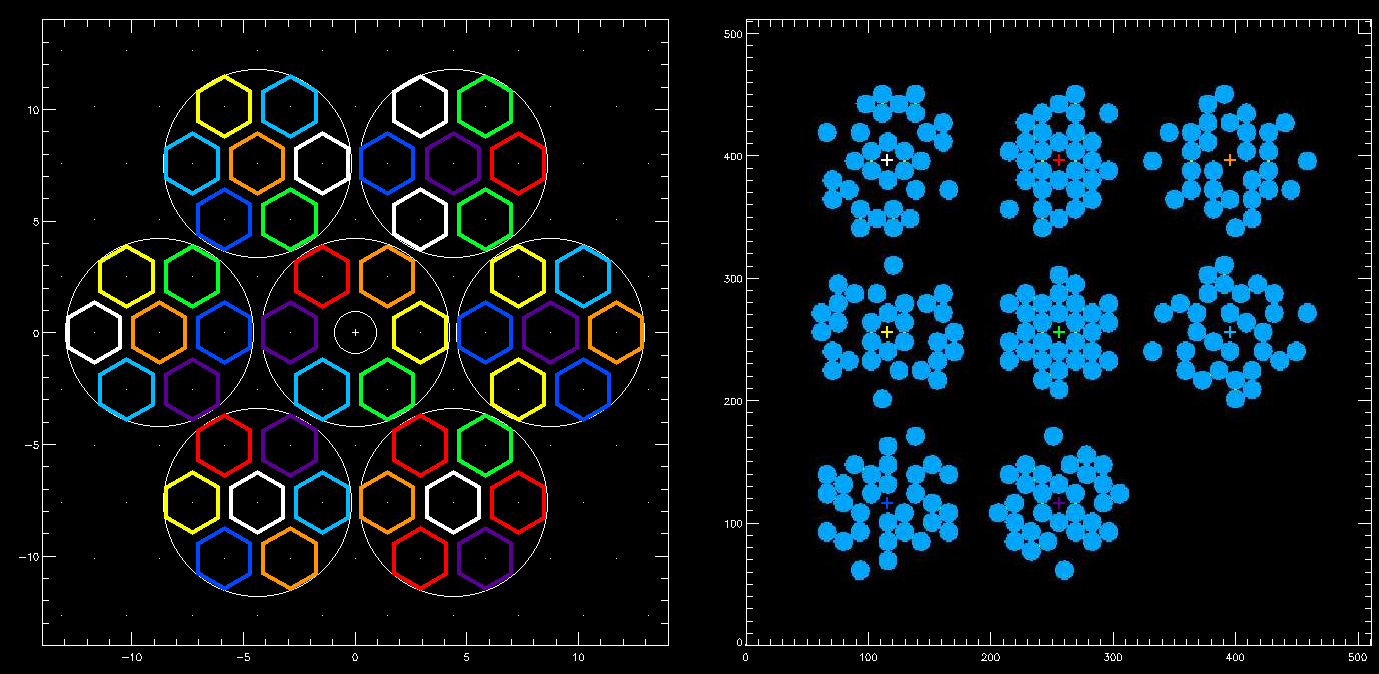}
    \caption{See caption for \autoref{fig:FC4x4s}.}
    \label{fig:GMT8x6s_48}
\end{figure}

\subsection*{Disclosures}
\noindent The authors declare no conflicts of interest.

\subsection* {Code, Data, and Materials Availability} 
The code and data used to generate the figures in the main body of this text are available at \href{https://github.com/ataras2/Jewel1}{this GitHub link \faGithubSquare}. 

\subsection* {Acknowledgements}

PT would like to thank Gerard van Belle for a conversation in Montreal from which the idea for Jewel Optics sprang. Max Charles and  Louis Desdoigts provided insightful discussions, as well as help in finalising the manuscript. We also thank the anonymous reviewers for helpful suggestions which improved the paper.
Microsoft Copilot was used in parts of the software development.

We would like to acknowledge the Gadigal People of the Eora nation, the traditional owners of the land on which most of this work was completed. 

We acknowledge support from Astralis -- Australia’s optical astronomy instrumentation Consortium -- through the Australian Government’s National Collaborative Research Infrastructure Strategy (NCRIS) Program. 
This work used the OptoFab and SA nodes of the NCRIS-enabled Australian National Fabrication Facility (ANFF).

An earlier sketch of this work was first published as a proceedings paper: A. Taras, G. Piroscia, P. G. Tuthill, “Jewel masks: non-redundant Fizeau beam combination without the guilt”, (Yokohama, Japan) (2024).


\bibliography{references, additional}   
\bibliographystyle{spiejour}   


\vspace{2ex}\noindent\textbf{Adam K. Taras}  is an optomechanical engineer at Astralis Sydney, designing and constructing key instruments in the proposed Asgard suite for the Very Large Telescope Interferometer (VLTI); namely Heimdallr, Baldr, Solarstein and Seidr. He graduated from The University of Sydney in 2023 with a B Engineering Honours (Mechatronics, Space major) and B Science (Physics). Adam's professional interests lie in the areas of instrumentation, innovative imaging techniques, and harnessing the power of forward modelling to optimise system design/data analysis. 

\vspace{2ex}\noindent\textbf{Grace Piroscia} is a first year PhD student at the University of Sydney, affiliated with the Sydney Institute for Astronomy. She graduated from the University of Sydney in 2023 with a Bachelor of Engineering Honours (Mechatronics) and a Bachelor of Science (Physics). Her research interests include differentiable optical modelling, high-resolution imagining techniques and astronomical instrumentation.


\end{spacing}
\end{document}